\def\templateIEEE{IEEE}
\def\templateICECET{ICECET}

\def\templateType{IEEE}

\ifx\templateType\templateArxiv
    \input{config/arxiv}
\else \ifx\templateType\templateIEEE
    \documentclass[
11pt,conference
]{IEEEtran}

\usepackage[table]{xcolor}

\usepackage[english]{babel}
\usepackage[utf8]{inputenc}

\usepackage{multicol}
\usepackage{afterpage}

\usepackage{wrapfig}

\usepackage{multirow}
\usepackage{makecell}
\usepackage{array}
\usepackage{rotating}
\usepackage{booktabs}
\usepackage[skip=3pt]{caption}
\usepackage{stfloats} 

\usepackage{cellspace} 
\usepackage{subcaption}
\usepackage{setspace}
\usepackage{placeins}

\usepackage{paralist}

\usepackage{amsmath,amssymb,amsfonts}
\usepackage{bm}

\usepackage{algorithm}\usepackage{algpseudocode}

\usepackage{expl3}

\usepackage{dirtytalk}
\usepackage{gensymb}

\usepackage{soul}
\usepackage[final]{pdfcomment}

 \usepackage{microtype}

\usepackage{hyperref}
\hypersetup{
    colorlinks,
    linkcolor={red!50!black},
    citecolor={blue!50!black},
    urlcolor={blue!80!black}
}

\newcolumntype{P}[1]{>{\centering\arraybackslash}p{#1}}

\newcolumntype{M}[1]{>{\centering\arraybackslash$}m{#1}<{$}}

\makeatletter
\def\refsetcounter{\@ifnextchar[{\refsetcounter@optarg}{\refsetcounter@noarg}}
\def\refsetcounter@noarg#1{\cref@constructprefix{#1}{\cref@result}\@ifundefined{cref@#1@alias}{\def\@tempa{#1}}{\def\@tempa{\csname cref@#1@alias\endcsname}}\protected@xdef\cref@currentlabel{[\@tempa][\arabic{#1}][\cref@result]\csname p@#1\endcsname\csname the#1\endcsname}\hyper@makecurrent{#1}}
\def\refsetcounter@optarg[#1]#2{\cref@constructprefix{#2}{\cref@result}\protected@xdef\cref@currentlabel{[#1][\arabic{#2}][\cref@result]\csname p@#2\endcsname\csname the#2\endcsname}\hyper@makecurrent{#2}}

\newcommand{\zref@uniqueautorefname}{AUTOzref}
\let\orig@caption=\caption
\def\caption#1{\orig@caption{#1}\refsetcounter{figure}}
\makeatother

\makeatletter
\newcommand{\linebreakand}{\end{@IEEEauthorhalign}
  \hfill\mbox{}\par
  \mbox{}\hfill\begin{@IEEEauthorhalign}
}
\makeatother

\usepackage{tikz}

\newcommand\copyrighttext{\footnotesize
  This work has been submitted to the IEEE for possible publication. Copyright may be transferred without notice, after which this version may no longer be accessible.
}
\newcommand\copyrightnotice{\begin{tikzpicture}[remember picture,overlay]
\node[anchor=south,yshift=10pt] at (current page.south) {\fbox{\parbox{\dimexpr\textwidth-\fboxsep-\fboxrule\relax}{\copyrighttext}}};
\end{tikzpicture}} 

 \else \ifx\templateType\templateElsevier
    \input{config/elsevier}
\else \ifx\templateType\templateICECET
    \documentclass[conference, Path=./ICECET_template/]{IEEEtran}

\makeatletter

\def\ps@IEEEtitlepagestyle{\def\@oddfoot{\mycopyrightnotice}\def\@evenfoot{}}
\def\mycopyrightnotice{{\footnotesize 979-8-3315-3559-9/25/\$31.00~\copyright~2025 IEEE\hfill}\gdef\mycopyrightnotice{}
}

\usepackage{blindtext}
\usepackage{eso-pic}
\IEEEoverridecommandlockouts
\usepackage{cite}
\usepackage{textcomp}

\usepackage[table]{xcolor}
\def\BibTeX{{\rm B\kern-.05em{\sc i\kern-.025em b}\kern-.08em
    T\kern-.1667em\lower.7ex\hbox{E}\kern-.125emX}}

\usepackage{eso-pic}
\newcommand\AtPageUpperMyright[1]{\AtPageUpperLeft{\put(\LenToUnit{0.17\paperwidth},\LenToUnit{-2cm}){\parbox{0.9\textwidth}{\raggedleft\fontsize{8}{11}\selectfont #1}}}}\newcommand{\conf}[1]{\AddToShipoutPictureBG*{\AtPageUpperMyright{#1}
}
}

\usepackage[english]{babel}
\usepackage[utf8]{inputenc}

\usepackage{multicol}
\usepackage{afterpage}

\usepackage{wrapfig}

\usepackage{multirow}
\usepackage{makecell}
\usepackage{array}
\usepackage{rotating}
\usepackage{booktabs}
\usepackage[skip=3pt]{caption}
\usepackage{stfloats} 

\usepackage{cellspace} 
\usepackage{subcaption}
\usepackage{setspace}
\usepackage{placeins}

\usepackage{paralist}

\usepackage{amsmath,amssymb,amsfonts}
\usepackage{bm}

\usepackage{algorithm}\usepackage{algpseudocode}

\usepackage{expl3}

\usepackage{dirtytalk}
\usepackage{gensymb}

\usepackage{soul}
\usepackage[final]{pdfcomment}

\usepackage{hyperref}
\hypersetup{
    colorlinks,
    linkcolor={blue!10!black},
    citecolor={blue!10!black},
    urlcolor={blue!10!black}
}

\newcolumntype{P}[1]{>{\centering\arraybackslash}p{#1}}

\newcolumntype{M}[1]{>{\centering\arraybackslash$}m{#1}<{$}}

\makeatletter
\def\refsetcounter{\@ifnextchar[{\refsetcounter@optarg}{\refsetcounter@noarg}}
\def\refsetcounter@noarg#1{\cref@constructprefix{#1}{\cref@result}\@ifundefined{cref@#1@alias}{\def\@tempa{#1}}{\def\@tempa{\csname cref@#1@alias\endcsname}}\protected@xdef\cref@currentlabel{[\@tempa][\arabic{#1}][\cref@result]\csname p@#1\endcsname\csname the#1\endcsname}\hyper@makecurrent{#1}}
\def\refsetcounter@optarg[#1]#2{\cref@constructprefix{#2}{\cref@result}\protected@xdef\cref@currentlabel{[#1][\arabic{#2}][\cref@result]\csname p@#2\endcsname\csname the#2\endcsname}\hyper@makecurrent{#2}}

\newcommand{\zref@uniqueautorefname}{AUTOzref}
\let\orig@caption=\caption
\def\caption#1{\orig@caption{#1}\refsetcounter{figure}}
\makeatother

 \fi \fi \fi \fi

\graphicspath{{./pics}{./pics/ds}{./pics/bsg}{./pics/split}{./pics/cmht_er}{./pics/pad_er}{./pics/pipe}}

\begin{document}

\ifx\templateType\templateIEEE
    
\title{Correcting Domain Shifts in Electric Motor Vibration Data for Unseen Operating Conditions
}

\author{
\IEEEauthorblockN{Lesley Wheat}
\IEEEauthorblockA{Department of Computing and Software,\\
McMaster University,
Hamilton, ON, Canada\\
Email: wheatd@mcmaster.ca}
\and
\IEEEauthorblockN{Martin v. Mohrenschildt}
\IEEEauthorblockA{Department of Computing and Software,\\
McMaster University,
Hamilton, ON, Canada}
\linebreakand
\IEEEauthorblockN{Saeid Habibi}
\IEEEauthorblockA{Center for Mechatronics and\\
Hybrid Technologies (CMHT),\\
Department of Mechanical Engineering,\\
McMaster University,
Hamilton, ON, Canada}
\and
\IEEEauthorblockN{Dhafar Al-Ani}
\IEEEauthorblockA{Department of Electrical and\\Computer Engineering,\\
McMaster University,
Hamilton, ON, Canada}
}

\maketitle
\copyrightnotice

\begin{abstract}
~{This paper addresses the problem of domain shifts in \pdfmarkupcomment[color=yellow]{electric}{Added.} motor vibration data created by new operating conditions in testing scenarios, focusing on bearing fault detection and diagnosis (FDD).
The proposed method combines the Harmonic Feature Space (HFS) with regression to correct for frequency and energy differentials in steady-state data, enabling accurate FDD on unseen operating conditions within the range of the training conditions.
The HFS aligns harmonics across different operating frequencies, while regression compensates for energy variations, preserving the relative magnitude of vibrations critical for fault detection.
The proposed approach is evaluated on a detection problem using experimental data from a Belt-Starter Generator (BSG) electric motor, with test conditions having a minimum 1000 RPM and 5 Nm difference from training conditions.
Results demonstrate that the method outperforms traditional analysis techniques, achieving high classification accuracy at a 94\% detection rate and effectively reducing domain shifts.
The approach is computationally efficient, requires only healthy data for training, and is well-suited for real-world applications where the exact application operating conditions cannot be predetermined.
 }
\end{abstract}

\begin{IEEEkeywords}
Fault Detection and Diagnosis,
Rolling Bearing,
Domain Adaptation,
Classification,
Vibration,
\pdfmarkupcomment[color=yellow]{Machine Learning}{
Keywords are only suggestions.
}
\end{IEEEkeywords}

 \else \ifx\templateType\templateICECET
    \title{\vspace*{1cm} Correcting Domain Shifts in Electric Motor Vibration Data for Unseen Operating Conditions\\
\thanks{
This research was supported by FedDev Ontario project 814996, Natural Sciences and Engineering Research Council of Canada (NSERC) Create project CREAT-482038-2016 and D\&V-NSERC Alliance project ALLRP-549016-2019.
}
}

\author{\IEEEauthorblockN{Lesley Wheat}
\IEEEauthorblockA{\textit{Department of} \\
\textit{Computing and Software} \\
\textit{McMaster University}\\
Hamilton, Canada\\
wheatd@mcmaster.ca}
\and
\IEEEauthorblockN{Martin v. Mohrenschildt}
\IEEEauthorblockA{\textit{Department of} \\
\textit{Computing and Software} \\
\textit{McMaster University}\\
Hamilton, Canada \\
mohrens@mcmaster.ca}
\and
\IEEEauthorblockN{Saeid Habibi}
\IEEEauthorblockA{\textit{Center for Mechatronics and} \\
\textit{Hybrid Technologies (CMHT)} \\
\textit{McMaster University}\\
Hamilton, Canada \\
habibi@mcmaster.ca}
\and
\IEEEauthorblockN{Dhafar Al-Ani}
\IEEEauthorblockA{\textit{Department of Electrical and} \\
\textit{Computer Engineering} \\
\textit{McMaster University}\\
Hamilton, Canada \\
mohammds@mcmaster.ca}
}

\maketitle
\conf{\textit{  V. International Conference on Electrical, Computer and Energy Technologies (ICECET 2025) \\ 
3-6 July 2025, Paris-France}}
\begin{abstract}
~{This paper addresses the problem of domain shifts in \pdfmarkupcomment[color=yellow]{electric}{Added.} motor vibration data created by new operating conditions in testing scenarios, focusing on bearing fault detection and diagnosis (FDD).
The proposed method combines the Harmonic Feature Space (HFS) with regression to correct for frequency and energy differentials in steady-state data, enabling accurate FDD on unseen operating conditions within the range of the training conditions.
The HFS aligns harmonics across different operating frequencies, while regression compensates for energy variations, preserving the relative magnitude of vibrations critical for fault detection.
The proposed approach is evaluated on a detection problem using experimental data from a Belt-Starter Generator (BSG) electric motor, with test conditions having a minimum 1000 RPM and 5 Nm difference from training conditions.
Results demonstrate that the method outperforms traditional analysis techniques, achieving high classification accuracy at a 94\% detection rate and effectively reducing domain shifts.
The approach is computationally efficient, requires only healthy data for training, and is well-suited for real-world applications where the exact application operating conditions cannot be predetermined.
 }
\end{abstract}

 \fi \fi 

\ifx\templateType\templateICECET
\begin{figure*}[tb]
\centering
\includegraphics[width=0.7\textwidth]{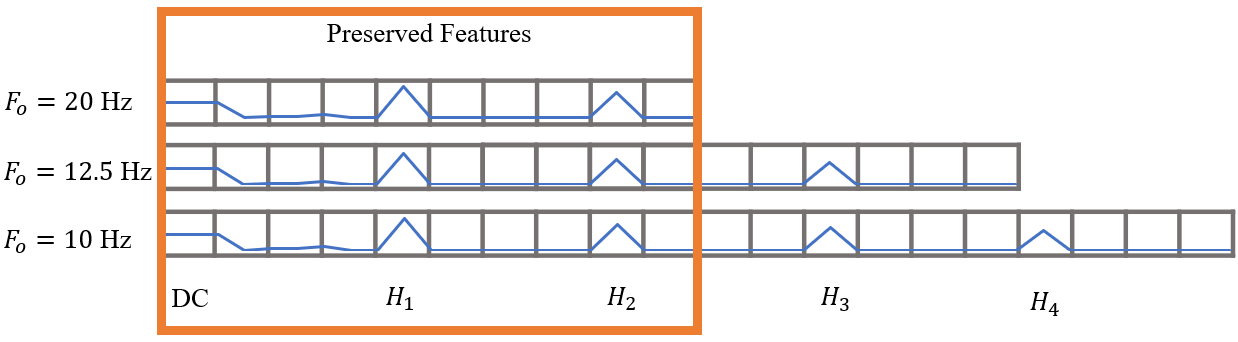}
\else
\begin{figure*}[b]
\centering
\includegraphics[width=0.99\textwidth]{harspace}
\fi

\caption{Example of a Harmonics Space Power Spectrum. $f_s=100$ Hz and $d=4$.}
\label{fig:harspace}
\end{figure*} \ifx\templateType\templateICECET
\begin{figure}[tb]
\else
\begin{figure*}[b]
\fi

\centering
\ifx\templateType\templateICECET
\begin{subfigure}{\linewidth}
\else
\begin{subfigure}{0.49\linewidth}
\fi
  \centering
  
  \ifx\templateType\templateICECET
  \includegraphics[width=0.8\linewidth]{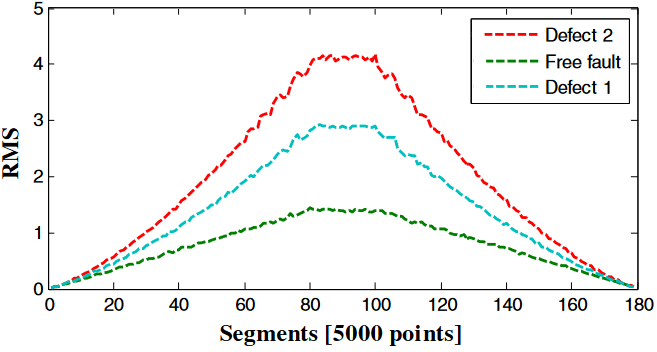}
  \else
  \includegraphics[width=0.99\linewidth]{bearingRampA.png}
  \fi
  
    \caption{Simulated RMS values for healthy bearing (Free fault) and two fault conditions.}
\end{subfigure}
\hfill
\ifx\templateType\templateICECET
\begin{subfigure}{\linewidth}
\else
\begin{subfigure}{0.47\linewidth}
\fi

  \centering
  \ifx\templateType\templateICECET
    \includegraphics[width=0.8\linewidth]{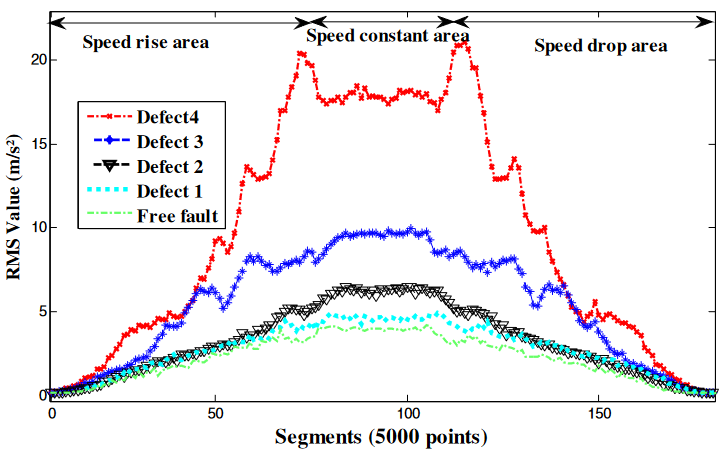}
  \else
    \includegraphics[width=0.99\linewidth]{real_rms.PNG}
  \fi
    \caption{Experimental data.}
\end{subfigure}

\caption{Examples of RMS data during ramp up to 950 RPM \cite{ait_sghir_bearing_2016}. Defect 1: 0.8 $\mbox{mm}^2$. Defect 2: 3.6 $\mbox{mm}^2$. Defect 3: 9.2 $\mbox{mm}^2$. Defect 4: 14 $\mbox{mm}^2$.}
\label{fig:ait}

\ifx\templateType\templateICECET
\end{figure}
\else
\end{figure*}
\fi 
\section{Introduction} \label{sec:intro}

Transferring techniques from controlled laboratory settings to real-world applications is a significant challenge in Fault Detection and Diagnosis (FDD) research, an area that seeks to detect and identify faults, undesirable conditions, within mechanical systems before failure occurs, which could result in expensive consequences.
One area that is facing this problems is bearing condition monitoring, a leading cause of \pdfmarkupcomment[color=yellow]{electric}{Added} motor failure \cite{bonnett_increased_2008}.
Automated condition monitoring could greatly benefit industries by improving preventative maintenance and reducing downtime, but faces several challenges in practical implementation.

In such case the goal is to use sensor data to determine if there is damage to the bearing, such as cracks or pitting.
It is known that these fault signatures are linked to the operating speed of the \pdfmarkupcomment[color=yellow]{electric}{Added} motor \cite{wong_transformer_based_2024}, and can be extracted in a model-based approach based on the specifications of the machinery.
When that information is not known, a data-driven approach must be taken, where the differences in the data created by the different operating conditions have not been accounted for.
If the model is trained and applied to the same condition (an assumption in much research \cite{wheat_impact_2024}), it does not have to learn to compensate for different conditions.
But, in many practical applications, the speed and load conditions under which the \pdfmarkupcomment[color=yellow]{electric}{Added} motor will operate can not be predetermined.

Thus, machine learning techniques must be adapted to learn to deal with reality, and operating condition based domain shifts.
A domain shift is a difference in class probability distributions when data is collected from different sources \cite{moreno-torres_unifying_2012}, which may cause problems in classification tasks, such as reduced performance \cite{moreno-torres_unifying_2012} or generalization issues.
\ifx\templateType\templateICECET
With this in mind, the objectives for the ideal technique are as follows:
\else
\noindent
\begin{minipage}{\linewidth}
\smallskip
With this in mind, the objectives for the ideal technique are as follows:
\fi
\begin{itemize}
    \item FDD: Allows for detection and diagnosis of bearing condition.
    \item \say{Blind} Approach: A data-driven method, or model-free such that the mechanical specifications are not required (such as bearing model, number of balls, etc).
    \item Automated: requires only sensor data (such as vibration or sound).
    \item Limited Information in Training: Speed and load are known at test time, but not present in the training data. Data from the bearing under test is not included in training \cite{wheat_impact_2024}.
    \item Desirable Characteristics: Explainability, robustness, reliability and generalizability.
\end{itemize}
\ifx\templateType\templateICECET\else
\end{minipage}
\medskip
\fi

Here, a new method is proposed for to adjust vibration data for unseen operating conditions that is based on a combination of the Harmonic Feature Space (HFS) \cite{hofer_model-free_2022} and regression.
By transforming the data into the HFS, the data is aligned across different operating frequencies, while the regression step compensates for the energy differential.

The effectiveness of the proposed approach is evaluated using experimental data from a Belt-Starter Generator (BSG) electric motor, which includes fifteen operating scenarios over 0-6000 RPM and 0-20 Nm.
The unseen operating conditions used for testing are: 5 Nm
at 2000 RPM, 5 Nm at 3000 RPM and 5 Nm at 4000 RPM.
Evaluation is done by using Principal Component Analysis (PCA) and k-Nearest Neighbor (k-NN) to measure detection accuracy and domain shifts.
The proposed method achieves a 94\% detection accuracy in testing, while frequency analysis only reaches 58 \% and envelope analysis 34\%.
When the domain shift is analyzed, it also shows a 38\% improvement when compared to frequency analysis.

\ifx\templateType\templateICECET
    \section{Background} \label{sec:bg}

While deep learning and neural networks have been a popular foundation of recent approaches for domain adaptation in vibration data with respect to differences in operating conditions \cite{yao_survey_2023, althobiani_novel_2024}.
Data processing has also been applied to this problem \cite{ait_sghir_bearing_2016, zhang_slice-oriented_2024}
.
Outside of domain adaptation, many different frequency extraction techniques have been used for rolling bearing fault detection and diagnosis including: Fast Fourier Transform (FFT), Short Time Fourier Transform (STFT), \pdfmarkupcomment[color=yellow]{Hilbert Fast Fourier Transform (HFFT),}{Added.} wavelets, envelope analysis and more \cite{wheat_impact_2024}.
Furthermore, work has been done to adjust RMS vibration values for a limited range of operating speeds to preform detection in transient conditions \cite{ait_sghir_bearing_2016}.

In \cite{ait_sghir_bearing_2016}, the authors make the assumption that the impulse response is speed invariant.
However, this only holds true for healthy bearings and limited operating speed ranges, for example ones that avoid resonance frequencies.
From \cite{ait_sghir_bearing_2016} (see \pdfmarkupcomment[color=yellow]{Fig.}{Template corrected} \ref{fig:ait}), some nonlinear behavior can be observed near the startup after correction, but otherwise, it can be observed that the RMS-speed relationship is near linear in the limited operating range (0-950 RPM).
The active power normalization shows promising results \cite{ait_sghir_bearing_2016}, although the transient RMS value for an operating speed may be larger than the steady-state because of the behavior of the \pdfmarkupcomment[color=yellow]{electric}{Added.} motor.

However, this technique only approaches the problem from an energy standpoint, when the different fault signatures are linked to different frequencies.
Therefore, RMS is only sufficient for detection.
To preform more complex diagnosis, more frequency information is required.
In order to preserve relative frequency information, the data can be transformed into the Harmonic Feature Space (HFS), an operating frequency independent domain \cite{hofer_model-free_2022}.
This is accomplished by adjusting the window size of the FFT, the harmonics become aligned, as shown in \pdfmarkupcomment[color=yellow]{Fig.}{Template corrected} \ref{fig:harspace}.

The window size $N$ is calculated using the following formula for sampling frequency $f_s$, operating frequency $f_o$ and harmonic bin $d$: 
\ifx\templateType\templateICECET
\pdfmarkupcomment[color=yellow]{$N = \text{round} ( f_s d / f_o )$.}{Shrunk}
\else
 \begin{equation}
     N = \text{round} \left( \frac{f_s d}{f_o} \right).
\end{equation}
\fi

In this case, a slight change to the method in \cite{hofer_model-free_2022} will be applied to account for the differences in energy due to the fault conditions, as this information needs to be preserved as well to enable detection and diagnosis.

     \ifx\templateType\templateICECET
    \begin{algorithm}[tb]
\else
    \begin{algorithm}[tbh]
\fi

 \caption{Creating Harmonic Hilbert Space} \label{alg:har}
 \hspace*{\algorithmicindent} \textbf{Input} $x$: signal vector\\
 \hspace*{\algorithmicindent} \textbf{Input} $d$: harmonic bin, $d \in \mathbb{N}$\\
 \hspace*{\algorithmicindent} \textbf{Input} $f_o$: operating frequency\\
 \hspace*{\algorithmicindent} \textbf{Input} $f_s$: sampling frequency \\
 \hspace*{\algorithmicindent} \textbf{Output} $H_{nm}$: feature matrix
 \ifx\templateType\templateICECET\else \\ \fi
 \begin{algorithmic}[1]
\State $N \gets $\Call{round}{$f_s * d / f_o$}
  \Comment{Calculate window size}
  \State $w \gets$ \Call{window$_{blackman}$}{$N$}
  \State $c \gets \sqrt{N/ \Call{sum}{w^2}}$
  \State $n \gets \Call{length}{x}$
  \For{$i = 1, 2, ..., \Call{floor}{n/N}$}
  \State $z \gets w * \Call{get\_segment}{z, i, N}$
  \Comment{Window Signal}
  \State $z \gets z - \Call{mean}{z}$
  \Comment{DC Filter}
  \State $h \gets$ \Call{envelope$_{hilbert}$}{$z$}
  \State $s \gets c * \Call{power\_spectrum}{h}$
  \State $H[i,:] \gets s/(N/2)$
  \EndFor
  \State \Return $H$
 \end{algorithmic}
 \end{algorithm}     \ifx\templateType\templateICECET
    \begin{algorithm}[tb]
\else
    \begin{algorithm}[tbh]
\fi
  
 \caption{Algorithm for Operating Condition Adjustment in Harmonic Space} \label{alg:adj}
  \hspace*{\algorithmicindent} \textbf{Input} $H_{nm}$: matrix with $n$ samples and $m$ features \\
  \hspace*{\algorithmicindent} \textbf{Input} $f_o$: vector of operating frequencies \\
  \hspace*{\algorithmicindent} \textbf{Input} $t_o$: vector of operating load \\
  \hspace*{\algorithmicindent} \textbf{Require} $f_o[i]$ and $t_o[i]$ are for $H[i,:]$ \\
  \hspace*{\algorithmicindent} \textbf{Input/Output} $A_{mp}$: model coefficients \\
  \hspace*{\algorithmicindent} \textbf{Output} $\hat{H}_{nm}$: adjusted values
   \ifx\templateType\templateICECET\else \\ \fi
   
 \begin{algorithmic}[1]
 \Statex \Call{make\_x}{$f_o, t_o$}
 \Comment{Create condition monomial matrix}
  \For{$i = 1, \dots, n$}
  \State $a \gets f_o[i]$
  \State $b  \gets t_o[i]$
  \State $X[i,:]  \gets [1,  a, b, a^2, ab, b^2]$
  \EndFor
  \Statex \Return $X$
  \Statex
   \end{algorithmic}
   \ifx\templateType\templateICECET \vspace{-0.5cm} \fi
\begin{algorithmic}[1]
 \Statex \Call{train}{$H_{nm},  f_o, t_o$}
 \State $X =$ \Call{make\_x}{$f_o, t_o$}
  \For{$i = 1, \dots, m$}
   \Comment{Fit for each feature}
  \State $y  \gets H[:,i]$
  \State $A[i,:]  \gets$ \Call{fit\_linear\_regression}{$X, y$}
  \State $A[i, 1]  \gets 0$ \Comment{Set offset to zero}
  \EndFor
  \Statex \Return $A$
\Statex
\end{algorithmic}
\ifx\templateType\templateICECET \vspace{-0.5cm} \fi
\begin{algorithmic}[1]
 \Statex \Call{adjust}{$H_{nm},  f_o, t_o, A_{mp}$}
 \State $X =$ \Call{make\_x}{$f_o, t_o$}
  \For{$i = 1, \dots, m$}
   \Comment{Adjust for each feature}
  \State $\hat{H}[:,i] \gets H[:,i] - A[i,:] * X[:, i] $
  \EndFor
  \Statex \Return $\hat{H}$
\end{algorithmic}
 \end{algorithm}     \section{Proposed Methodology} \label{sec:method}

In the application, speed changes in rotating machinery create two domain shifts: one related to the harmonics, and a second one related to the magnitude of the vibration energy. Due to the fact that the magnitude of the vibration is a key component for fault detection, the relative magnitude must be preserved. The power spectrum values are needed for diagnosis.

Transforming the data into the harmonic space \cite{hofer_model-free_2022} accounts for the variation in operating frequencies, but does not address the change in energy values.
Unlike as shown in \cite{hofer_model-free_2022}, samples can not be normalized by energy in this case.
Note that additional complexity is added because some energy is lost when samples are trimmed as shown in \pdfmarkupcomment[color=yellow]{Fig.}{Template corrected} \ref{fig:harspace}, and therefore lower operating frequencies contain less energy than expected.
Also, energy values can not simply be normalized per samples because the relative energy is an indicator of the fault condition.

While \cite{ait_sghir_bearing_2016} adjusts the RMS within a small operating speed range, the method does not meet the criteria for this application.
However, a similar foundation can be drawn.
From \pdfmarkupcomment[color=yellow]{Fig.}{Template corrected} \ref{fig:ait}, it can be seen that healthy and faulty bearings follow different models.
The state of the bearing under test in application can not be known in advance, therefore, this adjustment will be trained on healthy data and applied as if new data is healthy.
Even if the domain shift is solved within the healthy data, it is expected than they would still exist for the faults.

Based on the results of \cite{ait_sghir_bearing_2016}, it is expected that vibration energy to operating speed relationship is expected to be near-linear away from resonance frequencies.
As the energy is not evenly distributed across frequencies, a polynomial model of degree two can be fitted for each individual feature in the harmonics space.
A similar method was previously used as part of speed adjustment strategy in \cite{zhang_slice-oriented_2024}, but in this case, both speed and load are used as independent variables.

\ifx\templateType\templateICECET\else
\noindent 
\begin{minipage}{\linewidth}
\fi
Therefore, there are certain assumptions and limitations to bear in mind:
\begin{itemize}
\item All data is collected at steady state.
\item Unseen operating conditions must lie within the speed and load range of the training data.
\item None of the operating conditions are close to resonance points and the vibration data is near-linear in the selected speed range.
\item The state (heathy/faulty) of the training data is known.
\end{itemize}
\ifx\templateType\templateICECET\else
\smallskip
\end{minipage}
\fi

\ifx\templateType\templateICECET\else
\noindent 
\begin{minipage}{\linewidth} 
\fi
To adjust the data with the parameters of $f_o$ (operating frequency) and $T_o$ (operating load), the process is as follows:
\begin{enumerate}\item Measurement space: Collect healthy vibration signals for differing operating speeds \pdfmarkupcomment[color=yellow]{(RPM)}{added} and loads \pdfmarkupcomment[color=yellow]{(Nm)}{added} under steady state conditions.
\item Feature Extraction in Harmonic Space:
\begin{enumerate}
\item Apply Algorithm \ref{alg:har} to each signal. For multiple channels, the Euclidean magnitude over all channels at each frequency is used.
\item Given a maximum operating frequency, $N_{min}$ is the minimum window size. Keep $N_{min}/2$ features as shown in \pdfmarkupcomment[color=yellow]{Fig.}{Template corrected} \ref{fig:harspace}. Additional data is discarded so all operating speeds have the same number of features.
\item Processing: Filter DC and convert to decibels (dB).
\end{enumerate}
\item Train and apply operating condition adjustment \pdfmarkupcomment[color=yellow]{using ordinary least squares (ODS) linear regression with a polynomial of degree 2}{Added based on reviewer comment 1.3} to the resulting feature matrix (Algorithm \ref{alg:adj}).
\end{enumerate}
\ifx\templateType\templateICECET\else
\smallskip
\end{minipage}
\fi

 \else
    \bigskip
    \section{Background} \label{sec:bg}

While deep learning and neural networks have been a popular foundation of recent approaches for domain adaptation in vibration data with respect to differences in operating conditions \cite{yao_survey_2023, althobiani_novel_2024}.
Data processing has also been applied to this problem \cite{ait_sghir_bearing_2016, zhang_slice-oriented_2024}
.
Outside of domain adaptation, many different frequency extraction techniques have been used for rolling bearing fault detection and diagnosis including: Fast Fourier Transform (FFT), Short Time Fourier Transform (STFT), \pdfmarkupcomment[color=yellow]{Hilbert Fast Fourier Transform (HFFT),}{Added.} wavelets, envelope analysis and more \cite{wheat_impact_2024}.
Furthermore, work has been done to adjust RMS vibration values for a limited range of operating speeds to preform detection in transient conditions \cite{ait_sghir_bearing_2016}.

In \cite{ait_sghir_bearing_2016}, the authors make the assumption that the impulse response is speed invariant.
However, this only holds true for healthy bearings and limited operating speed ranges, for example ones that avoid resonance frequencies.
From \cite{ait_sghir_bearing_2016} (see \pdfmarkupcomment[color=yellow]{Fig.}{Template corrected} \ref{fig:ait}), some nonlinear behavior can be observed near the startup after correction, but otherwise, it can be observed that the RMS-speed relationship is near linear in the limited operating range (0-950 RPM).
The active power normalization shows promising results \cite{ait_sghir_bearing_2016}, although the transient RMS value for an operating speed may be larger than the steady-state because of the behavior of the \pdfmarkupcomment[color=yellow]{electric}{Added.} motor.

However, this technique only approaches the problem from an energy standpoint, when the different fault signatures are linked to different frequencies.
Therefore, RMS is only sufficient for detection.
To preform more complex diagnosis, more frequency information is required.
In order to preserve relative frequency information, the data can be transformed into the Harmonic Feature Space (HFS), an operating frequency independent domain \cite{hofer_model-free_2022}.
This is accomplished by adjusting the window size of the FFT, the harmonics become aligned, as shown in \pdfmarkupcomment[color=yellow]{Fig.}{Template corrected} \ref{fig:harspace}.

The window size $N$ is calculated using the following formula for sampling frequency $f_s$, operating frequency $f_o$ and harmonic bin $d$: 
\ifx\templateType\templateICECET
\pdfmarkupcomment[color=yellow]{$N = \text{round} ( f_s d / f_o )$.}{Shrunk}
\else
 \begin{equation}
     N = \text{round} \left( \frac{f_s d}{f_o} \right).
\end{equation}
\fi

In this case, a slight change to the method in \cite{hofer_model-free_2022} will be applied to account for the differences in energy due to the fault conditions, as this information needs to be preserved as well to enable detection and diagnosis.

     \section{Proposed Methodology} \label{sec:method}

In the application, speed changes in rotating machinery create two domain shifts: one related to the harmonics, and a second one related to the magnitude of the vibration energy. Due to the fact that the magnitude of the vibration is a key component for fault detection, the relative magnitude must be preserved. The power spectrum values are needed for diagnosis.

Transforming the data into the harmonic space \cite{hofer_model-free_2022} accounts for the variation in operating frequencies, but does not address the change in energy values.
Unlike as shown in \cite{hofer_model-free_2022}, samples can not be normalized by energy in this case.
Note that additional complexity is added because some energy is lost when samples are trimmed as shown in \pdfmarkupcomment[color=yellow]{Fig.}{Template corrected} \ref{fig:harspace}, and therefore lower operating frequencies contain less energy than expected.
Also, energy values can not simply be normalized per samples because the relative energy is an indicator of the fault condition.

While \cite{ait_sghir_bearing_2016} adjusts the RMS within a small operating speed range, the method does not meet the criteria for this application.
However, a similar foundation can be drawn.
From \pdfmarkupcomment[color=yellow]{Fig.}{Template corrected} \ref{fig:ait}, it can be seen that healthy and faulty bearings follow different models.
The state of the bearing under test in application can not be known in advance, therefore, this adjustment will be trained on healthy data and applied as if new data is healthy.
Even if the domain shift is solved within the healthy data, it is expected than they would still exist for the faults.

Based on the results of \cite{ait_sghir_bearing_2016}, it is expected that vibration energy to operating speed relationship is expected to be near-linear away from resonance frequencies.
As the energy is not evenly distributed across frequencies, a polynomial model of degree two can be fitted for each individual feature in the harmonics space.
A similar method was previously used as part of speed adjustment strategy in \cite{zhang_slice-oriented_2024}, but in this case, both speed and load are used as independent variables.

\ifx\templateType\templateICECET\else
\noindent 
\begin{minipage}{\linewidth}
\fi
Therefore, there are certain assumptions and limitations to bear in mind:
\begin{itemize}
\item All data is collected at steady state.
\item Unseen operating conditions must lie within the speed and load range of the training data.
\item None of the operating conditions are close to resonance points and the vibration data is near-linear in the selected speed range.
\item The state (heathy/faulty) of the training data is known.
\end{itemize}
\ifx\templateType\templateICECET\else
\smallskip
\end{minipage}
\fi

\ifx\templateType\templateICECET\else
\noindent 
\begin{minipage}{\linewidth} 
\fi
To adjust the data with the parameters of $f_o$ (operating frequency) and $T_o$ (operating load), the process is as follows:
\begin{enumerate}\item Measurement space: Collect healthy vibration signals for differing operating speeds \pdfmarkupcomment[color=yellow]{(RPM)}{added} and loads \pdfmarkupcomment[color=yellow]{(Nm)}{added} under steady state conditions.
\item Feature Extraction in Harmonic Space:
\begin{enumerate}
\item Apply Algorithm \ref{alg:har} to each signal. For multiple channels, the Euclidean magnitude over all channels at each frequency is used.
\item Given a maximum operating frequency, $N_{min}$ is the minimum window size. Keep $N_{min}/2$ features as shown in \pdfmarkupcomment[color=yellow]{Fig.}{Template corrected} \ref{fig:harspace}. Additional data is discarded so all operating speeds have the same number of features.
\item Processing: Filter DC and convert to decibels (dB).
\end{enumerate}
\item Train and apply operating condition adjustment \pdfmarkupcomment[color=yellow]{using ordinary least squares (ODS) linear regression with a polynomial of degree 2}{Added based on reviewer comment 1.3} to the resulting feature matrix (Algorithm \ref{alg:adj}).
\end{enumerate}
\ifx\templateType\templateICECET\else
\smallskip
\end{minipage}
\fi

     \ifx\templateType\templateICECET
    \begin{algorithm}[tb]
\else
    \begin{algorithm}[tbh]
\fi

 \caption{Creating Harmonic Hilbert Space} \label{alg:har}
 \hspace*{\algorithmicindent} \textbf{Input} $x$: signal vector\\
 \hspace*{\algorithmicindent} \textbf{Input} $d$: harmonic bin, $d \in \mathbb{N}$\\
 \hspace*{\algorithmicindent} \textbf{Input} $f_o$: operating frequency\\
 \hspace*{\algorithmicindent} \textbf{Input} $f_s$: sampling frequency \\
 \hspace*{\algorithmicindent} \textbf{Output} $H_{nm}$: feature matrix
 \ifx\templateType\templateICECET\else \\ \fi
 \begin{algorithmic}[1]
\State $N \gets $\Call{round}{$f_s * d / f_o$}
  \Comment{Calculate window size}
  \State $w \gets$ \Call{window$_{blackman}$}{$N$}
  \State $c \gets \sqrt{N/ \Call{sum}{w^2}}$
  \State $n \gets \Call{length}{x}$
  \For{$i = 1, 2, ..., \Call{floor}{n/N}$}
  \State $z \gets w * \Call{get\_segment}{z, i, N}$
  \Comment{Window Signal}
  \State $z \gets z - \Call{mean}{z}$
  \Comment{DC Filter}
  \State $h \gets$ \Call{envelope$_{hilbert}$}{$z$}
  \State $s \gets c * \Call{power\_spectrum}{h}$
  \State $H[i,:] \gets s/(N/2)$
  \EndFor
  \State \Return $H$
 \end{algorithmic}
 \end{algorithm}     \ifx\templateType\templateICECET
    \begin{algorithm}[tb]
\else
    \begin{algorithm}[tbh]
\fi
  
 \caption{Algorithm for Operating Condition Adjustment in Harmonic Space} \label{alg:adj}
  \hspace*{\algorithmicindent} \textbf{Input} $H_{nm}$: matrix with $n$ samples and $m$ features \\
  \hspace*{\algorithmicindent} \textbf{Input} $f_o$: vector of operating frequencies \\
  \hspace*{\algorithmicindent} \textbf{Input} $t_o$: vector of operating load \\
  \hspace*{\algorithmicindent} \textbf{Require} $f_o[i]$ and $t_o[i]$ are for $H[i,:]$ \\
  \hspace*{\algorithmicindent} \textbf{Input/Output} $A_{mp}$: model coefficients \\
  \hspace*{\algorithmicindent} \textbf{Output} $\hat{H}_{nm}$: adjusted values
   \ifx\templateType\templateICECET\else \\ \fi
   
 \begin{algorithmic}[1]
 \Statex \Call{make\_x}{$f_o, t_o$}
 \Comment{Create condition monomial matrix}
  \For{$i = 1, \dots, n$}
  \State $a \gets f_o[i]$
  \State $b  \gets t_o[i]$
  \State $X[i,:]  \gets [1,  a, b, a^2, ab, b^2]$
  \EndFor
  \Statex \Return $X$
  \Statex
   \end{algorithmic}
   \ifx\templateType\templateICECET \vspace{-0.5cm} \fi
\begin{algorithmic}[1]
 \Statex \Call{train}{$H_{nm},  f_o, t_o$}
 \State $X =$ \Call{make\_x}{$f_o, t_o$}
  \For{$i = 1, \dots, m$}
   \Comment{Fit for each feature}
  \State $y  \gets H[:,i]$
  \State $A[i,:]  \gets$ \Call{fit\_linear\_regression}{$X, y$}
  \State $A[i, 1]  \gets 0$ \Comment{Set offset to zero}
  \EndFor
  \Statex \Return $A$
\Statex
\end{algorithmic}
\ifx\templateType\templateICECET \vspace{-0.5cm} \fi
\begin{algorithmic}[1]
 \Statex \Call{adjust}{$H_{nm},  f_o, t_o, A_{mp}$}
 \State $X =$ \Call{make\_x}{$f_o, t_o$}
  \For{$i = 1, \dots, m$}
   \Comment{Adjust for each feature}
  \State $\hat{H}[:,i] \gets H[:,i] - A[i,:] * X[:, i] $
  \EndFor
  \Statex \Return $\hat{H}$
\end{algorithmic}
 \end{algorithm} \fi

\ifx\templateType\templateICECET
    \ifx\templateType\templateICECET
\begin{figure}[t]
\else
\begin{figure}[h]
\fi

\begin{subfigure}{\linewidth}
  \centering
  \includegraphics[width=0.8\linewidth]{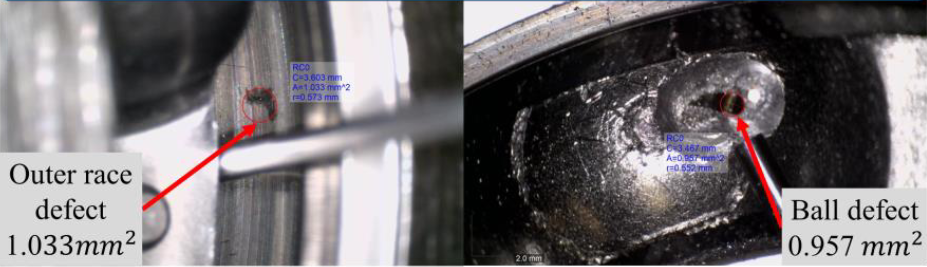}
    \caption{Close up photos of damage to F3-01.}
    \label{fig:f3}
\end{subfigure}
\hfill
\begin{subfigure}{\linewidth}
  \centering
  \includegraphics[width=0.8\linewidth]{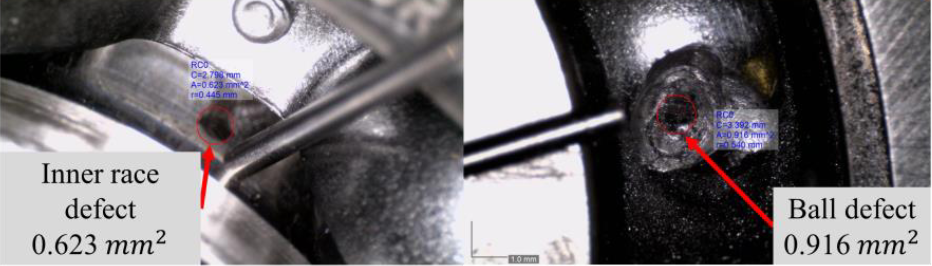}
    \caption{Close up photos of damage to F5-01.}
    \label{fig:f5}
\end{subfigure}
\hfill
\begin{subfigure}{\linewidth}
  \centering
  \includegraphics[width=0.8\linewidth]{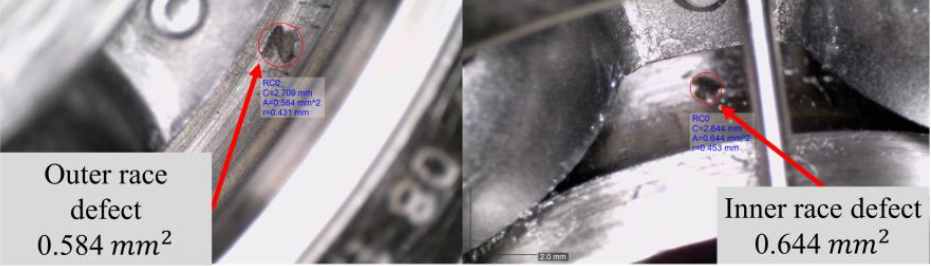}
    \caption{Close up photos of damage to F7-01.}
    \label{fig:f7}
\end{subfigure}
\hfill
\begin{subfigure}{0.99\linewidth}
\centering
\includegraphics[width=0.4\linewidth, trim={10cm 4cm 3cm 0},clip]{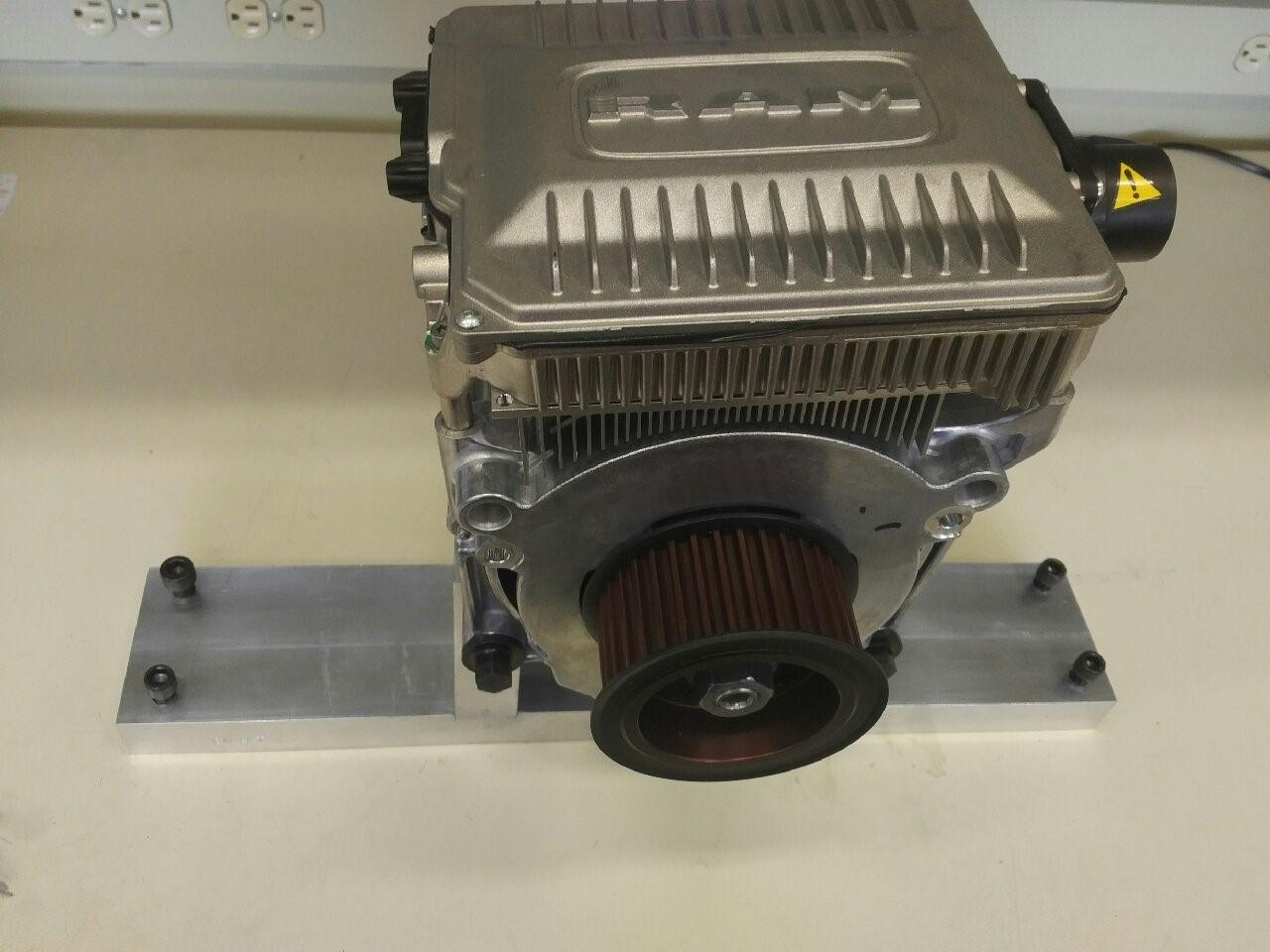}
\includegraphics[width=0.4\linewidth, trim={1cm 0.5cm 0.5cm 3.4cm},clip]{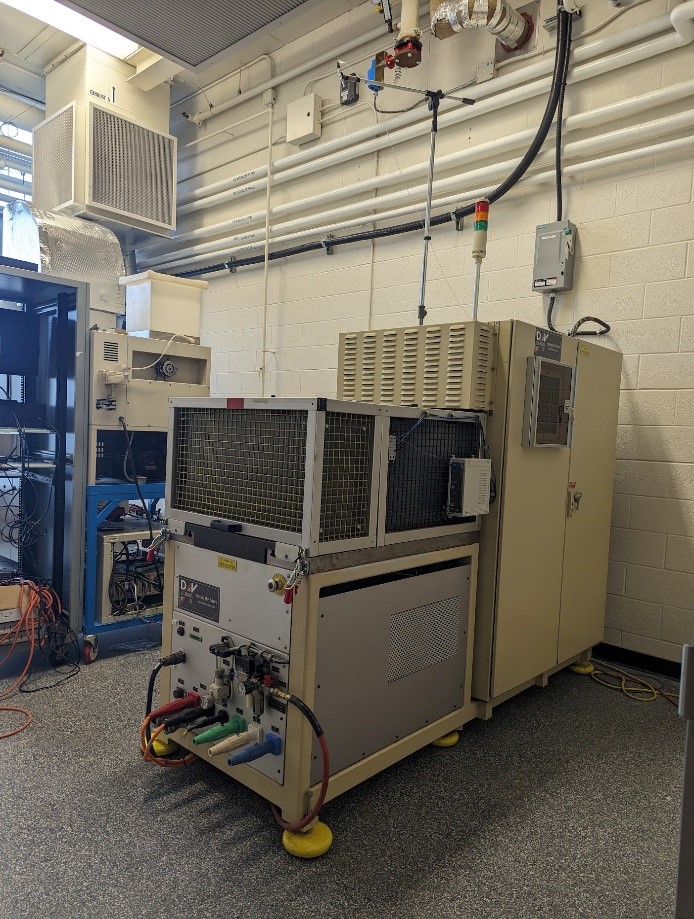}
    \caption{BSG (left), and HT-250 test bench (right).}
    \label{fig:bsgpics}
\end{subfigure}
\hfill
\begin{subfigure}{0.99\linewidth}
  \centering
  \includegraphics[width=0.8\linewidth]{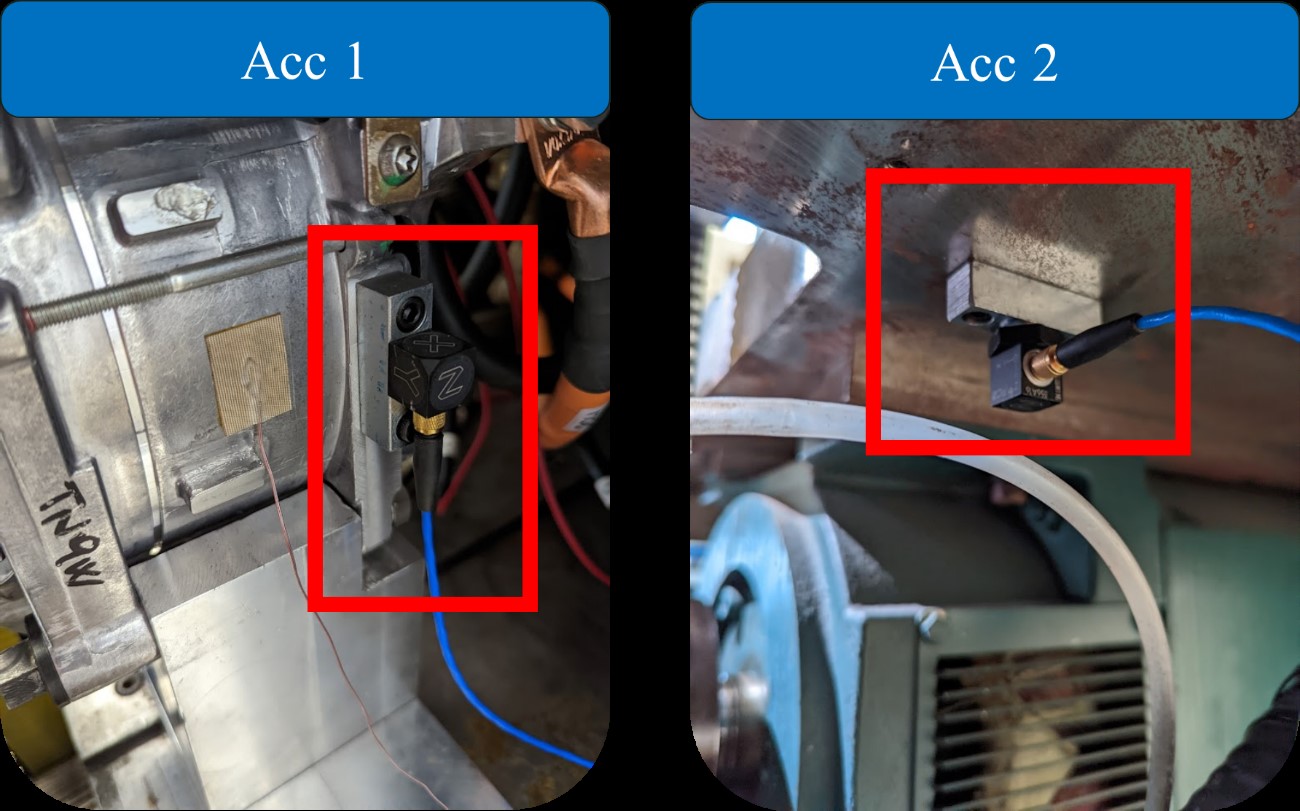}
    \caption{Location of accelerometers on motor.\\ 
    Accelerometer 1 ($A_1$): Right side of the BSG's chassis.\\
    Accelerometer 2 ($A_2$): Underneath the test bench.}
    \label{fig:acc}
\end{subfigure}

\caption{CMHT BSG Test Setup \cite{vazquez_technical_unpub, wong_transformer_based_2024} .}
\label{fig:cmht_bsg}
\end{figure}     \ifx\templateType\templateICECET
\begin{table}[t]
\else
\begin{table*}[tb]
\fi
\centering

\caption{Experimental data used from the CMHT Dataset \cite{wong_transformer_based_2024, vazquez_technical_unpub}.
The area of the respective hole is listed next to the defect type 
\ifx\templateType\templateIEEE
(see Fig. \ref{fig:closeup}).
\else
(see Fig. \ref{fig:cmht_bsg}).
\fi
The bearing model number is SKF 6305-2RS1/C3 \cite{vazquez_technical_unpub}.}
\label{table:cmht_bf}

\ifx\templateType\templateICECET
\begin{tabular*}{\linewidth}{@{\extracolsep{\fill}} 
P{0.8cm}P{2.1cm}P{2.2cm}P{1.2cm}}\hline
\else
\begin{tabular*}{\linewidth}{@{\extracolsep{\fill}} 
P{0.8cm}P{4cm}P{4cm}P{3cm}}\hline
\fi

Class & Type & Bearing \pdfmarkupcomment[color=yellow]{Identifier (ID)}{Added.} & Number of Runs\\
\hline

\multirow{3}{*}{Healthy}&\multirow{3}{2cm}{After Market} & \multirow{1}{*}{AM-01} & 6 \\  \cline{3-4}
&& AM-02 & 2\\ \cline{3-4}
&& AM-03 & 2\\ \cline{3-4}

\hline

\multirow{6}{*}{Faulty}
&OR (1.0 $\mbox{mm}^2$)&\multirow{2}{*}{F3-01}&\multirow{2}{*}{4}\\ 
&BD (1.0 $\mbox{mm}^2$)&&\\  \cline{2-4}

&IR (0.6 $\mbox{mm}^2$)&\multirow{2}{*}{F5-01}&\multirow{2}{*}{4}\\
&BD (0.9 $\mbox{mm}^2$)&&\\ \cline{2-4}

&IR (0.6 $\mbox{mm}^2$)&\multirow{2}{*}{F7-01}&\multirow{2}{*}{4}\\ 
&OR (0.6 $\mbox{mm}^2$)&&\\  \cline{2-4}
\hline

\multicolumn{4}{l}{\makecell[l]{
OR: Outer Ring; IR: Inner Ring; BD: Ball Defect;
}} \\ \hline
\end{tabular*}

\ifx\templateType\templateICECET
\end{table}
\else
\end{table*}
\fi     \ifx\templateType\templateICECET
\begin{figure}[t]
\else
\begin{figure}[h]
\fi

\centering

\ifx\templateType\templateICECET
\includegraphics[width=0.9\linewidth]{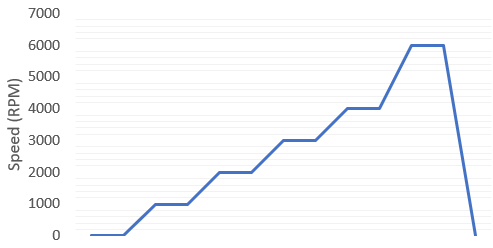}
\else
\includegraphics[width=0.99\linewidth]{ramp.PNG}
\fi

\caption{Diagram of Ramp Up Profile from 0 RPM to 6000 RPM. Steady state periods are held for approximately one second.}
\label{fig:speed_profile}
\end{figure} 
\section{Dataset} \label{sec:cmht}

Bearing fault vibration data from a Belt-Starter Generator (BSG) electric motor (Model BM1048564F) using a testbench (Model HT-250, D\&V Electronics LTD.) was obtained from CMHT \cite{vazquez_technical_unpub}.
Vibration data from two tri-axial accelerometers (Model No. 356A16, PCB Piezotronics, Inc.) were used, as shown in \pdfmarkupcomment[color=yellow]{Fig.}{Template corrected} \ref{fig:acc}, and all channels were sampled at 48 kHz \cite{vazquez_technical_unpub}.
More detailed information in provided in \cite{wong_transformer_based_2024}.

The dataset includes several combinations of artificial fault conditions for the drive-end bearing over twenty different speed and load conditions for the purpose of evaluating FDD techniques on an industry-typical scenario.
In this case, a subset of the dataset is used (shown in Table \ref{table:cmht_bf}), which excludes all lubrication fault conditions, operating speeds over 6000 RPM and loads over 20 Nm.
\pdfmarkupcomment[color=yellow]{Only one model of bearing is used.}{Moved text to table.}

In total, fifteen operating conditions were used.
To collect data for different speeds, the \pdfmarkupcomment[color=yellow]{electric}{Added.} motor speed was changed multiple times during the test (See Fig. \ref{fig:speed_profile}) and the continuous steady-state segments (approx. one second in length) were extracted into separate files.
Every file is pre-processed with a low pass Butterworth zero phase digital filter before feature extraction.

     \ifx\templateType\templateICECET
\begin{figure*}[tb]
\centering
\includegraphics[width=0.6\linewidth]{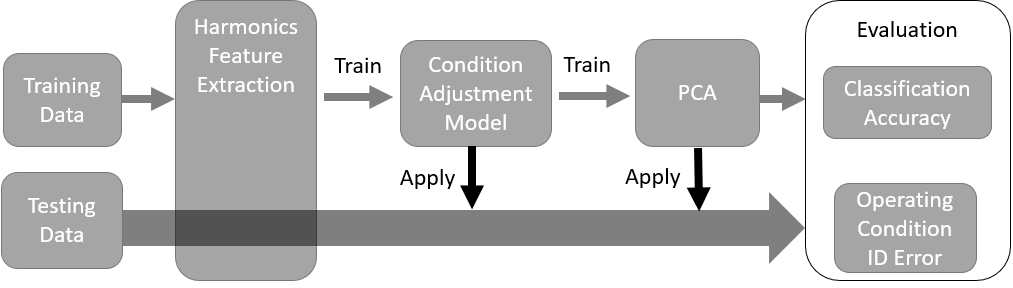}
\else
\begin{figure*}[tb]
\centering
\includegraphics[width=0.8\linewidth]{evalpipe.png}
\fi

\caption{HARH learning and evaluation pipeline.}
\label{fig:pipelines}
\end{figure*}     \ifx\templateType\templateICECET
\begin{figure}[tb]
\else
\begin{figure*}[ht]
\fi

\centering

\ifx\templateType\templateICECET
\begin{subfigure}{0.64\linewidth}
\else
\begin{subfigure}{0.36\linewidth}
\fi

  \centering
  \includegraphics[width=1\linewidth]{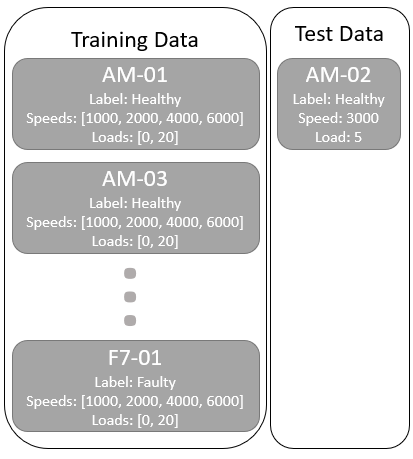}
    \caption{Split for Classification Accuracy.}
    \label{fig:split_acc}
\end{subfigure}
\ifx\templateType\templateICECET
\begin{subfigure}{0.34\linewidth}
\else
\begin{subfigure}{0.19\linewidth}
\fi

  \centering
  \includegraphics[width=0.65\linewidth]{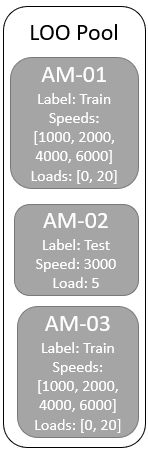}
    \caption{Split for Operating Condition ID Error.}
    \label{fig:split_ocid}
\end{subfigure}

\caption{Example Train-Test Splits for Test Set: AM-02.}
\label{fig:split}

\ifx\templateType\templateICECET
\end{figure}
\else
\end{figure*}
\fi

\section{Experiments}

The scenario to be tested involves determining whether a new bearing under a previously unseen operating condition can be accurately classified.
To avoid the issue of overfitting to specific bearings \cite{wheat_impact_2024}, each bearing is also considered to be a different domain for evaluation purposes.
As such, data from each bearing only exists within the training or testing dataset, not both.
Therefore, the test data must be unique on the basis of: bearing ID, speed and load (see \pdfmarkupcomment[color=yellow]{Fig.}{Template corrected} \ref{fig:split}).
Keeping with interpolation only, the unseen operating conditions are: 5 Nm at 2000 RPM, 5 Nm at 3000 RPM and 5 Nm at \pdfmarkupcomment[color=yellow]{4000 RPM.}{There is a new line here. Line spacing is controlled by the template and can't be changed.}

Note that when fault conditions are used for testing, this means that no data on that particular fault conditions exists within the training data.
However, as the fault conditions are combinations (Table \ref{table:cmht_bf}), elements of those conditions are still present in the training data.

\pdfmarkupcomment[color=yellow]{After}{Removed \ noindent.} splitting the dataset, the processing method is applied.
In order to compare against HARH from Section \ref{sec:method}, FFT and HFFT are also applied, two feature extraction techniques previously used on the CMHT dataset \cite{wheat_impact_2024}.
In total, four different feature extraction methods were used \pdfmarkupcomment[color=yellow]{and named as follows}{Clarified.}:

\ifx\templateType\templateICECET\else
\noindent
\begin{minipage}{\linewidth}
\smallskip
\fi
\begin{itemize}
  \item \pdfmarkupcomment[color=yellow]{FFT}{FFT is defined in Section 2: Background.}: Frequency Analysis \cite{wheat_impact_2024}
  \item \pdfmarkupcomment[color=yellow]{HFFT}{Added to Section 2: Background.}: Envelope Analysis \cite{wheat_impact_2024}
  \item \pdfmarkupcomment[color=yellow]{HARH}{Not an acronym.}: Hilbert Harmonics (see \pdfmarkupcomment[color=yellow]{Fig.}{Template corrected} \ref{fig:pipelines})
  \item \pdfmarkupcomment[color=yellow]{HAR}{Not an acronym.}: Harmonics (HARH without Hilbert Transform)
\end{itemize}
\ifx\templateType\templateICECET\else
\smallskip
\end{minipage}
\fi

\ifx\templateType\templateICECET\else
\noindent
\begin{minipage}{\linewidth}
\fi
\noindent Details of the process for FFT and HFFT:
\begin{enumerate}\item Extract features (window size: 8196)
\item Remove DC
\item Low pass filter at 6000 Hz
\end{enumerate}
\ifx\templateType\templateICECET\else
\smallskip \end{minipage} \fi

\ifx\templateType\templateICECET\else
\noindent
\begin{minipage}{\linewidth}
\fi
\noindent Details of the process for HAR and HARH:
\begin{enumerate}\item Train and apply adjustment (Section \ref{sec:method}) (bin: 4)
\item Low pass filter at 60 harmonics
\end{enumerate}
\ifx\templateType\templateICECET\else
\smallskip
\end{minipage}
\fi

In order to test the effectiveness of the data processing, a standard classification strategy is used, to avoid any chance of the classifier compensating for any remaining domain shift.
To this end, a combination of Principal Component Analysis (PCA) and k-Nearest Neighbor (kNN) are applied.
PCA is used as a dimensionality reduction strategy, as kNN is susceptible to the curse of dimensionality.

To search for the optimal value k* for kNN, the kNN classifier is run over the training data using a \pdfmarkupcomment[color=yellow]{leave-one-out (LOO)}{Added.} split for a range of k values, then the k with the lowest error (reweighted to account for class imbalance) is taken as k*.
The minimum value of k is 1 and the maximum value is limited to the bearing in the training set with the least number of samples.
The output of kNN(k*) classifier applied to the test set given the training data is taken as the classification accuracy (see \pdfmarkupcomment[color=yellow]{Fig.}{Template corrected} \ref{fig:split_acc}).

In the case two sets of data have the same probability distributions, it should not be possible to separate the two.
Thus, the amount of domain shift can be measured by how easy it is to classify the two sets of data.
kNN is a good choice for this, as it has a long history in being used to measure differences in probability distributions \cite{cover_nearest_1967}.
To this effect, the classifier can be applied to the same data, but with different class labels, to check for domain shifts.

First, the two classes are expected to have different probability distributions, so only data from the same class can be compared.
If the test set is a healthy bearing, then all the healthy data from the training set is extracted and the data is pooled together with the class labels becoming \say{train} and \say{test}.
On that pool of data kNN(k*) is applied with a LOO strategy and the classification error of the test data becomes the operating condition ID error (see \pdfmarkupcomment[color=yellow]{Fig.}{Template corrected} \ref{fig:split_ocid}).
\pdfmarkupcomment[color=yellow]{kNN is preferred for this evaluation over two sample tests, which look for statistically significant differences in distributions, because kNN measures the \textit{scale} of the difference (rather than it's existence) and is less dependant on the total number of samples (set by the window size, in this case).
}{Added based on reviewer comment 1.4}
\pdfmarkupcomment[color=yellow]{
A low error indicates the data is very separable and has a large domain shift. 
}{Removed previous sentence}

\ifx\templateType\templateICECET\else
\noindent 
\begin{minipage}{\linewidth}
\medskip
\fi
The experimental process is summarized as follows:
\begin{enumerate}\item Split dataset into train and test sets
\item Processing: Apply FFT, HFFT, HAR or HARH.
\item Dimensionality Reduction: Scale all remaining features to have zero mean and unit variance then apply PCA (preserved features: 2).
\item Evaluation:
\begin{enumerate}
\item Classification Accuracy: kNN(k*) accuracy on test set.
\item Set Identification Error: kNN(k*) LOO error on same class with \say{train} and \say{test} as the labels.
\end{enumerate}
\end{enumerate}
\ifx\templateType\templateICECET\else
\smallskip
\end{minipage}
\fi

\ifx\templateType\templateICECET
\begin{table}[tb]
\else
\begin{table*}[thb]
\fi

\centering
\caption{HARH with $A_1 + A_2$.}

\begin{subtable}{1\linewidth}
\centering
\caption{Classification Accuracy.}
\label{table:results_a12_acc}
\begin{tabular}{ P{0.5cm}cccccc } 
\hline
&AM-01&AM-02&AM-03&F3-01&F5-01&F7-01\\
\hline

2000&100\%& 100\%& 100\%& 100\%& 100\%& 100\% \\
3000&0\%& 100\%& 100\%& 100\%& 100\%& 100\% \\
4000&100\%& 100\%& 100\%& 100\%& 100\%& 100\% \\

\hline

\end{tabular}
\end{subtable}
\begin{subtable}{1\linewidth}
\ifx\templateType\templateICECET\else\smallskip\fi
\centering
\caption{Operating Condition ID Error (higher is better).}
\label{table:results_a12_opcond}
\begin{tabular}{ P{0.5cm}cccccc } 
\hline
&AM-01&AM-02&AM-03&F3-01&F5-01&F7-01\\
\hline

2000&78\%&100\%& 50\%& 83\%& 100\%& 100\%\\
3000&6\%& 61\%& 44\%& 42\%& 86\%& 64\%\\
4000&35\%& 77\%& 81\%& 40\%& 50\%& 75\%\\

\hline

\end{tabular}
\end{subtable}

\ifx\templateType\templateICECET
\end{table}
\else
\end{table*}
\fi     \ifx\templateType\templateICECET
\begin{table}[tb]
\else
\begin{table}[h]
\fi
\centering
\caption{Results for different methods and different sources.}

\begin{subtable}{1\linewidth}
\centering
\caption{Classification accuracy.}
\label{table:results_acc}
\begin{tabular}{ ccccc } 
\hline
& FFT & HFFT & HAR & HARH\\
\hline

$A_1$&59\% & 34\%& 88\%& 89\% \\
$A_2$&31\%& 30\%& 50\%& 63\% \\
$A_1 + A_2$& 58\%& 34\%& 80\%& 94\% \\

\hline

\end{tabular}
\end{subtable}

\begin{subtable}{1\linewidth}
\centering
\ifx\templateType\templateICECET\else\smallskip\fi
\caption{Operating condition ID error (higher is better).}
\label{table:results_opcond}
\begin{tabular}{ ccccc } 
\hline
& FFT & HFFT & HAR & HARH\\
\hline

$A_1$& 23\%& 33\%& 34\%& 59\% \\
$A_2$& 12\%& 13\%& 53\%& 39\% \\
$A_1 + A_2$& 30\%& 41\%& 33\%& 68\% \\

\hline

\end{tabular}
\end{subtable}

\ifx\templateType\templateICECET
\end{table}
\else
\end{table}
\fi     \section{Results} \label{sec:results}

The classification accuracy for each process is reported in Table \ref{table:results_acc} and the operating condition ID error is reported in Table \ref{table:results_opcond}.
As there is an imbalance in the data available for the individual bearings (see Table \ref{table:cmht_bf}), the results are reweighted so every bearing is weighted equally.

From Table \ref{table:results_acc}, the best classification accuracy is 94\% with both accelerometers using HARH.
Upon further examination (Table \ref{table:results_a12_acc}), Bearing AM-01 at 3000 RPM is the only case that is incorrectly classified and only has a 6\% operating condition ID error, indicating that that condition is easily differentiated from the others and likely follows a different distribution.
However, as the two other healthy bearings, AM-02 and AM-03 have good results, it may be an outlier case.

Overall, the HARH method shows promise for the intended application, within the limitations outlined in Section \ref{sec:method}.
Note that all test conditions in this experiment have a minimum 1000 RPM difference from the training conditions, however the step size and range of the ramp up profile will likely require modification for different \pdfmarkupcomment[color=yellow]{electric}{added} motors (due to type, application, resonances, etc.).
In addition to being a fast and simple method, another advantage of HARH is that it requires only healthy data for training, which is usually easier and much less expensive to collect for many applications.

Assuming the results of the domain shift analysis on PCA-reduced data (Table \ref{table:results_a12_opcond}) apply to the original HARH space, the domain shift would also be corrected within that space.
Therefore, not just detection, but also diagnosis ought to be possible within the HARH space, as the fault signatures are linked to the relative operating frequencies of the \pdfmarkupcomment[color=yellow]{electric}{added} motor.
However, properly testing this hypothesis would require more faulty bearings for testing than are currently available in the dataset.

\else
    \ifx\templateType\templateICECET
\begin{table}[t]
\else
\begin{table*}[tb]
\fi
\centering

\caption{Experimental data used from the CMHT Dataset \cite{wong_transformer_based_2024, vazquez_technical_unpub}.
The area of the respective hole is listed next to the defect type 
\ifx\templateType\templateIEEE
(see Fig. \ref{fig:closeup}).
\else
(see Fig. \ref{fig:cmht_bsg}).
\fi
The bearing model number is SKF 6305-2RS1/C3 \cite{vazquez_technical_unpub}.}
\label{table:cmht_bf}

\ifx\templateType\templateICECET
\begin{tabular*}{\linewidth}{@{\extracolsep{\fill}} 
P{0.8cm}P{2.1cm}P{2.2cm}P{1.2cm}}\hline
\else
\begin{tabular*}{\linewidth}{@{\extracolsep{\fill}} 
P{0.8cm}P{4cm}P{4cm}P{3cm}}\hline
\fi

Class & Type & Bearing \pdfmarkupcomment[color=yellow]{Identifier (ID)}{Added.} & Number of Runs\\
\hline

\multirow{3}{*}{Healthy}&\multirow{3}{2cm}{After Market} & \multirow{1}{*}{AM-01} & 6 \\  \cline{3-4}
&& AM-02 & 2\\ \cline{3-4}
&& AM-03 & 2\\ \cline{3-4}

\hline

\multirow{6}{*}{Faulty}
&OR (1.0 $\mbox{mm}^2$)&\multirow{2}{*}{F3-01}&\multirow{2}{*}{4}\\ 
&BD (1.0 $\mbox{mm}^2$)&&\\  \cline{2-4}

&IR (0.6 $\mbox{mm}^2$)&\multirow{2}{*}{F5-01}&\multirow{2}{*}{4}\\
&BD (0.9 $\mbox{mm}^2$)&&\\ \cline{2-4}

&IR (0.6 $\mbox{mm}^2$)&\multirow{2}{*}{F7-01}&\multirow{2}{*}{4}\\ 
&OR (0.6 $\mbox{mm}^2$)&&\\  \cline{2-4}
\hline

\multicolumn{4}{l}{\makecell[l]{
OR: Outer Ring; IR: Inner Ring; BD: Ball Defect;
}} \\ \hline
\end{tabular*}

\ifx\templateType\templateICECET
\end{table}
\else
\end{table*}
\fi     \ifx\templateType\templateICECET
\begin{figure}[tb]
\else
\begin{figure}[ht]
\fi

\begin{subfigure}{0.99\linewidth}
\centering
\includegraphics[width=0.45\linewidth, trim={10cm 4cm 3cm 0},clip]{Figure3_A1.jpg}
\includegraphics[width=0.45\linewidth, trim={1cm 0.5cm 0.5cm 3.4cm},clip]{Figure3_A2.jpg}
    \caption{BSG (left), and HT-250 test bench (right).}
    \label{fig:bsgpics}
\end{subfigure}
\hfill
\begin{subfigure}{0.99\linewidth}
  \centering
  \includegraphics[width=0.9\linewidth]{Figure3_B.jpg}
    \caption{Location of accelerometers on motor.\\ 
    Accelerometer 1 ($A_1$): Right side of the BSG's chassis.\\
    Accelerometer 2 ($A_2$): Underneath the test bench.}
    \label{fig:acc}
\end{subfigure}

\caption{CMHT BSG Test Setup \cite{vazquez_technical_unpub, wong_transformer_based_2024} .}
\label{fig:cmht_bsg}
\end{figure}     \ifx\templateType\templateICECET
\begin{figure}[tb]
\else
\begin{figure}[h]
\fi

\begin{subfigure}{\linewidth}
  \centering
  \includegraphics[width=0.99\linewidth]{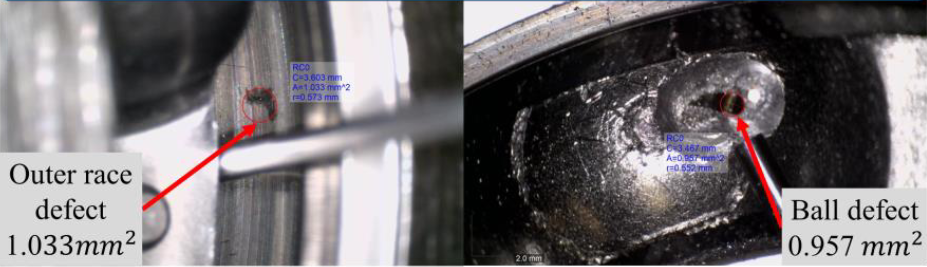}
    \caption{Close up photos of damage to F3-01.}
    \label{fig:f3}
\end{subfigure}
\hfill
\begin{subfigure}{\linewidth}
  \centering
  \includegraphics[width=0.99\linewidth]{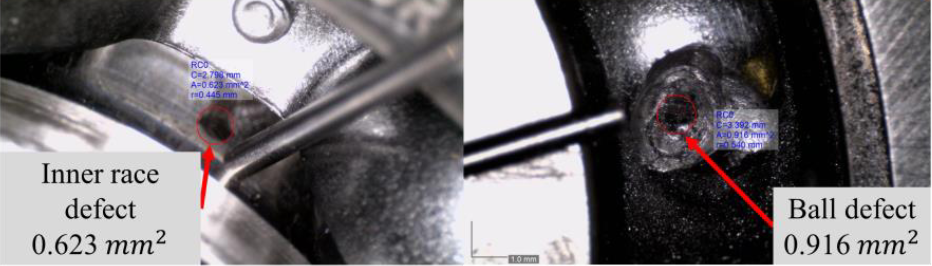}
    \caption{Close up photos of damage to F5-01.}
    \label{fig:f5}
\end{subfigure}
\hfill
\begin{subfigure}{\linewidth}
  \centering
  \includegraphics[width=0.99\linewidth]{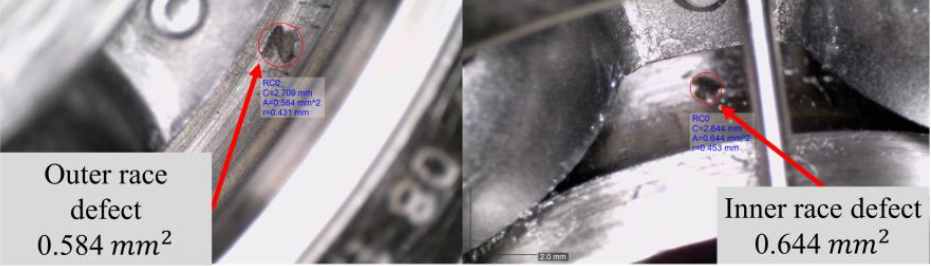}
    \caption{Close up photos of damage to F7-01.}
    \label{fig:f7}
\end{subfigure}
\caption{CMHT BSG Bearing Faults \cite{vazquez_technical_unpub, wong_transformer_based_2024}.}
\label{fig:closeup}
\end{figure}     \ifx\templateType\templateICECET
\begin{figure*}[tb]
\centering
\includegraphics[width=0.6\linewidth]{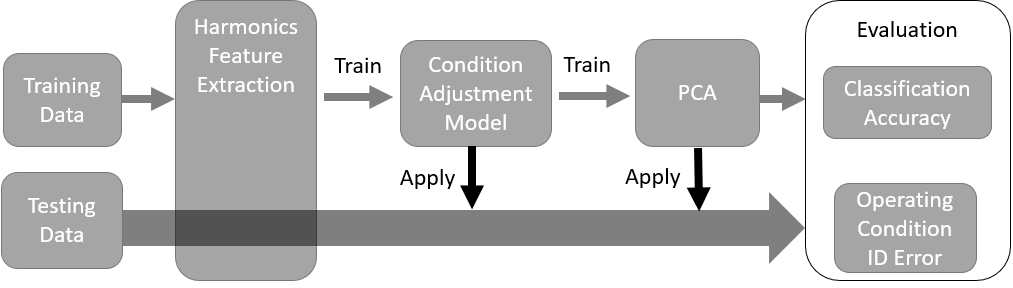}
\else
\begin{figure*}[tb]
\centering
\includegraphics[width=0.8\linewidth]{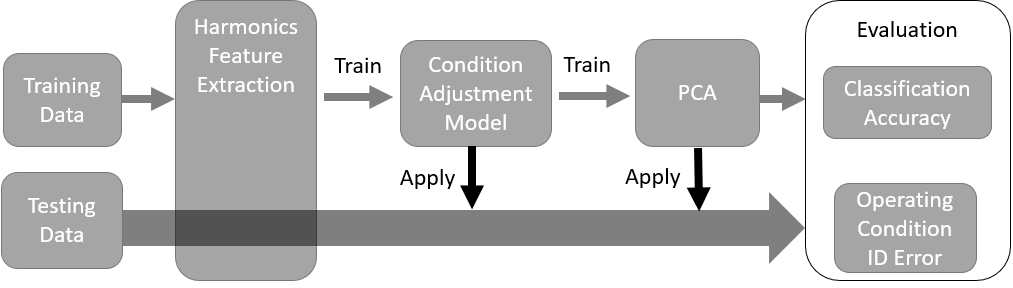}
\fi

\caption{HARH learning and evaluation pipeline.}
\label{fig:pipelines}
\end{figure*}     \ifx\templateType\templateICECET
\begin{figure}[t]
\else
\begin{figure}[h]
\fi

\centering

\ifx\templateType\templateICECET
\includegraphics[width=0.9\linewidth]{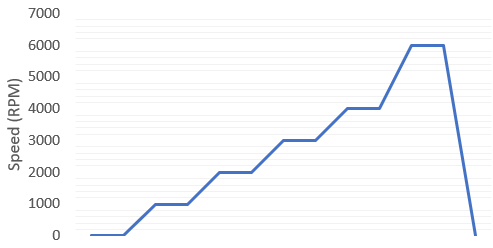}
\else
\includegraphics[width=0.99\linewidth]{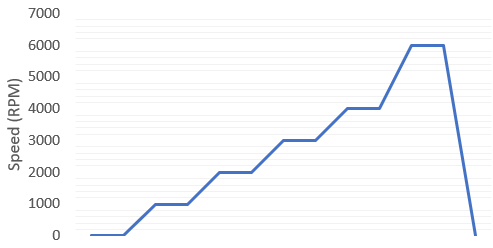}
\fi

\caption{Diagram of Ramp Up Profile from 0 RPM to 6000 RPM. Steady state periods are held for approximately one second.}
\label{fig:speed_profile}
\end{figure} 
\section{Dataset} \label{sec:cmht}

Bearing fault vibration data from a Belt-Starter Generator (BSG) electric motor (Model BM1048564F) using a testbench (Model HT-250, D\&V Electronics LTD.) was obtained from CMHT \cite{vazquez_technical_unpub}.
Vibration data from two tri-axial accelerometers (Model No. 356A16, PCB Piezotronics, Inc.) were used, as shown in \pdfmarkupcomment[color=yellow]{Fig.}{Template corrected} \ref{fig:acc}, and all channels were sampled at 48 kHz \cite{vazquez_technical_unpub}.
More detailed information in provided in \cite{wong_transformer_based_2024}.

The dataset includes several combinations of artificial fault conditions for the drive-end bearing over twenty different speed and load conditions for the purpose of evaluating FDD techniques on an industry-typical scenario.
In this case, a subset of the dataset is used (shown in Table \ref{table:cmht_bf}), which excludes all lubrication fault conditions, operating speeds over 6000 RPM and loads over 20 Nm.
\pdfmarkupcomment[color=yellow]{Only one model of bearing is used.}{Moved text to table.}

In total, fifteen operating conditions were used.
To collect data for different speeds, the \pdfmarkupcomment[color=yellow]{electric}{Added.} motor speed was changed multiple times during the test (See Fig. \ref{fig:speed_profile}) and the continuous steady-state segments (approx. one second in length) were extracted into separate files.
Every file is pre-processed with a low pass Butterworth zero phase digital filter before feature extraction.

     \ifx\templateType\templateICECET
\begin{figure}[tb]
\else
\begin{figure*}[ht]
\fi

\centering

\ifx\templateType\templateICECET
\begin{subfigure}{0.64\linewidth}
\else
\begin{subfigure}{0.36\linewidth}
\fi

  \centering
  \includegraphics[width=1\linewidth]{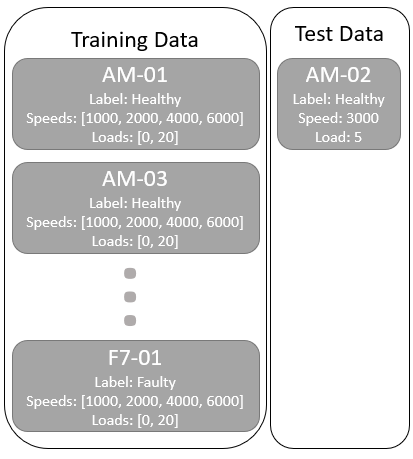}
    \caption{Split for Classification Accuracy.}
    \label{fig:split_acc}
\end{subfigure}
\ifx\templateType\templateICECET
\begin{subfigure}{0.34\linewidth}
\else
\begin{subfigure}{0.19\linewidth}
\fi

  \centering
  \includegraphics[width=0.65\linewidth]{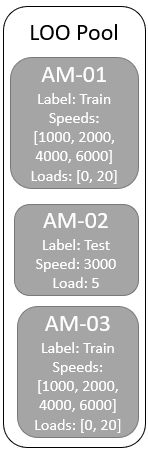}
    \caption{Split for Operating Condition ID Error.}
    \label{fig:split_ocid}
\end{subfigure}

\caption{Example Train-Test Splits for Test Set: AM-02.}
\label{fig:split}

\ifx\templateType\templateICECET
\end{figure}
\else
\end{figure*}
\fi

\section{Experiments}

The scenario to be tested involves determining whether a new bearing under a previously unseen operating condition can be accurately classified.
To avoid the issue of overfitting to specific bearings \cite{wheat_impact_2024}, each bearing is also considered to be a different domain for evaluation purposes.
As such, data from each bearing only exists within the training or testing dataset, not both.
Therefore, the test data must be unique on the basis of: bearing ID, speed and load (see \pdfmarkupcomment[color=yellow]{Fig.}{Template corrected} \ref{fig:split}).
Keeping with interpolation only, the unseen operating conditions are: 5 Nm at 2000 RPM, 5 Nm at 3000 RPM and 5 Nm at \pdfmarkupcomment[color=yellow]{4000 RPM.}{There is a new line here. Line spacing is controlled by the template and can't be changed.}

Note that when fault conditions are used for testing, this means that no data on that particular fault conditions exists within the training data.
However, as the fault conditions are combinations (Table \ref{table:cmht_bf}), elements of those conditions are still present in the training data.

\pdfmarkupcomment[color=yellow]{After}{Removed \ noindent.} splitting the dataset, the processing method is applied.
In order to compare against HARH from Section \ref{sec:method}, FFT and HFFT are also applied, two feature extraction techniques previously used on the CMHT dataset \cite{wheat_impact_2024}.
In total, four different feature extraction methods were used \pdfmarkupcomment[color=yellow]{and named as follows}{Clarified.}:

\ifx\templateType\templateICECET\else
\noindent
\begin{minipage}{\linewidth}
\smallskip
\fi
\begin{itemize}
  \item \pdfmarkupcomment[color=yellow]{FFT}{FFT is defined in Section 2: Background.}: Frequency Analysis \cite{wheat_impact_2024}
  \item \pdfmarkupcomment[color=yellow]{HFFT}{Added to Section 2: Background.}: Envelope Analysis \cite{wheat_impact_2024}
  \item \pdfmarkupcomment[color=yellow]{HARH}{Not an acronym.}: Hilbert Harmonics (see \pdfmarkupcomment[color=yellow]{Fig.}{Template corrected} \ref{fig:pipelines})
  \item \pdfmarkupcomment[color=yellow]{HAR}{Not an acronym.}: Harmonics (HARH without Hilbert Transform)
\end{itemize}
\ifx\templateType\templateICECET\else
\smallskip
\end{minipage}
\fi

\ifx\templateType\templateICECET\else
\noindent
\begin{minipage}{\linewidth}
\fi
\noindent Details of the process for FFT and HFFT:
\begin{enumerate}\item Extract features (window size: 8196)
\item Remove DC
\item Low pass filter at 6000 Hz
\end{enumerate}
\ifx\templateType\templateICECET\else
\smallskip \end{minipage} \fi

\ifx\templateType\templateICECET\else
\noindent
\begin{minipage}{\linewidth}
\fi
\noindent Details of the process for HAR and HARH:
\begin{enumerate}\item Train and apply adjustment (Section \ref{sec:method}) (bin: 4)
\item Low pass filter at 60 harmonics
\end{enumerate}
\ifx\templateType\templateICECET\else
\smallskip
\end{minipage}
\fi

In order to test the effectiveness of the data processing, a standard classification strategy is used, to avoid any chance of the classifier compensating for any remaining domain shift.
To this end, a combination of Principal Component Analysis (PCA) and k-Nearest Neighbor (kNN) are applied.
PCA is used as a dimensionality reduction strategy, as kNN is susceptible to the curse of dimensionality.

To search for the optimal value k* for kNN, the kNN classifier is run over the training data using a \pdfmarkupcomment[color=yellow]{leave-one-out (LOO)}{Added.} split for a range of k values, then the k with the lowest error (reweighted to account for class imbalance) is taken as k*.
The minimum value of k is 1 and the maximum value is limited to the bearing in the training set with the least number of samples.
The output of kNN(k*) classifier applied to the test set given the training data is taken as the classification accuracy (see \pdfmarkupcomment[color=yellow]{Fig.}{Template corrected} \ref{fig:split_acc}).

In the case two sets of data have the same probability distributions, it should not be possible to separate the two.
Thus, the amount of domain shift can be measured by how easy it is to classify the two sets of data.
kNN is a good choice for this, as it has a long history in being used to measure differences in probability distributions \cite{cover_nearest_1967}.
To this effect, the classifier can be applied to the same data, but with different class labels, to check for domain shifts.

First, the two classes are expected to have different probability distributions, so only data from the same class can be compared.
If the test set is a healthy bearing, then all the healthy data from the training set is extracted and the data is pooled together with the class labels becoming \say{train} and \say{test}.
On that pool of data kNN(k*) is applied with a LOO strategy and the classification error of the test data becomes the operating condition ID error (see \pdfmarkupcomment[color=yellow]{Fig.}{Template corrected} \ref{fig:split_ocid}).
\pdfmarkupcomment[color=yellow]{kNN is preferred for this evaluation over two sample tests, which look for statistically significant differences in distributions, because kNN measures the \textit{scale} of the difference (rather than it's existence) and is less dependant on the total number of samples (set by the window size, in this case).
}{Added based on reviewer comment 1.4}
\pdfmarkupcomment[color=yellow]{
A low error indicates the data is very separable and has a large domain shift. 
}{Removed previous sentence}

\ifx\templateType\templateICECET\else
\noindent 
\begin{minipage}{\linewidth}
\medskip
\fi
The experimental process is summarized as follows:
\begin{enumerate}\item Split dataset into train and test sets
\item Processing: Apply FFT, HFFT, HAR or HARH.
\item Dimensionality Reduction: Scale all remaining features to have zero mean and unit variance then apply PCA (preserved features: 2).
\item Evaluation:
\begin{enumerate}
\item Classification Accuracy: kNN(k*) accuracy on test set.
\item Set Identification Error: kNN(k*) LOO error on same class with \say{train} and \say{test} as the labels.
\end{enumerate}
\end{enumerate}
\ifx\templateType\templateICECET\else
\smallskip
\end{minipage}
\fi

\ifx\templateType\templateICECET
\begin{table}[tb]
\else
\begin{table}[h]
\fi
\centering
\caption{Results for different methods and different sources.}

\begin{subtable}{1\linewidth}
\centering
\caption{Classification accuracy.}
\label{table:results_acc}
\begin{tabular}{ ccccc } 
\hline
& FFT & HFFT & HAR & HARH\\
\hline

$A_1$&59\% & 34\%& 88\%& 89\% \\
$A_2$&31\%& 30\%& 50\%& 63\% \\
$A_1 + A_2$& 58\%& 34\%& 80\%& 94\% \\

\hline

\end{tabular}
\end{subtable}

\begin{subtable}{1\linewidth}
\centering
\ifx\templateType\templateICECET\else\smallskip\fi
\caption{Operating condition ID error (higher is better).}
\label{table:results_opcond}
\begin{tabular}{ ccccc } 
\hline
& FFT & HFFT & HAR & HARH\\
\hline

$A_1$& 23\%& 33\%& 34\%& 59\% \\
$A_2$& 12\%& 13\%& 53\%& 39\% \\
$A_1 + A_2$& 30\%& 41\%& 33\%& 68\% \\

\hline

\end{tabular}
\end{subtable}

\ifx\templateType\templateICECET
\end{table}
\else
\end{table}
\fi     \ifx\templateType\templateICECET
\begin{table}[tb]
\else
\begin{table*}[thb]
\fi

\centering
\caption{HARH with $A_1 + A_2$.}

\begin{subtable}{1\linewidth}
\centering
\caption{Classification Accuracy.}
\label{table:results_a12_acc}
\begin{tabular}{ P{0.5cm}cccccc } 
\hline
&AM-01&AM-02&AM-03&F3-01&F5-01&F7-01\\
\hline

2000&100\%& 100\%& 100\%& 100\%& 100\%& 100\% \\
3000&0\%& 100\%& 100\%& 100\%& 100\%& 100\% \\
4000&100\%& 100\%& 100\%& 100\%& 100\%& 100\% \\

\hline

\end{tabular}
\end{subtable}
\begin{subtable}{1\linewidth}
\ifx\templateType\templateICECET\else\smallskip\fi
\centering
\caption{Operating Condition ID Error (higher is better).}
\label{table:results_a12_opcond}
\begin{tabular}{ P{0.5cm}cccccc } 
\hline
&AM-01&AM-02&AM-03&F3-01&F5-01&F7-01\\
\hline

2000&78\%&100\%& 50\%& 83\%& 100\%& 100\%\\
3000&6\%& 61\%& 44\%& 42\%& 86\%& 64\%\\
4000&35\%& 77\%& 81\%& 40\%& 50\%& 75\%\\

\hline

\end{tabular}
\end{subtable}

\ifx\templateType\templateICECET
\end{table}
\else
\end{table*}
\fi     \section{Results} \label{sec:results}

The classification accuracy for each process is reported in Table \ref{table:results_acc} and the operating condition ID error is reported in Table \ref{table:results_opcond}.
As there is an imbalance in the data available for the individual bearings (see Table \ref{table:cmht_bf}), the results are reweighted so every bearing is weighted equally.

From Table \ref{table:results_acc}, the best classification accuracy is 94\% with both accelerometers using HARH.
Upon further examination (Table \ref{table:results_a12_acc}), Bearing AM-01 at 3000 RPM is the only case that is incorrectly classified and only has a 6\% operating condition ID error, indicating that that condition is easily differentiated from the others and likely follows a different distribution.
However, as the two other healthy bearings, AM-02 and AM-03 have good results, it may be an outlier case.

Overall, the HARH method shows promise for the intended application, within the limitations outlined in Section \ref{sec:method}.
Note that all test conditions in this experiment have a minimum 1000 RPM difference from the training conditions, however the step size and range of the ramp up profile will likely require modification for different \pdfmarkupcomment[color=yellow]{electric}{added} motors (due to type, application, resonances, etc.).
In addition to being a fast and simple method, another advantage of HARH is that it requires only healthy data for training, which is usually easier and much less expensive to collect for many applications.

Assuming the results of the domain shift analysis on PCA-reduced data (Table \ref{table:results_a12_opcond}) apply to the original HARH space, the domain shift would also be corrected within that space.
Therefore, not just detection, but also diagnosis ought to be possible within the HARH space, as the fault signatures are linked to the relative operating frequencies of the \pdfmarkupcomment[color=yellow]{electric}{added} motor.
However, properly testing this hypothesis would require more faulty bearings for testing than are currently available in the dataset.

 \fi

\section{Conclusion}

The proposed novel method (HARH) \pdfmarkupcomment[color=yellow]{based on Hilbert Harmonics,}{Added.} outperforms all other tested analysis techniques, for both detection and domain adaptation metrics, on the CMHT BSG dataset.
It is also practical, and can easily make use of existing preset profiles from other \pdfmarkupcomment[color=yellow]{tasks}{shortened}, as well as not requiring much training faulty data for detection, as this can be difficult to obtain in practice.
While the results are very promising, further testing with a larger dataset, including more fault conditions \pdfmarkupcomment[color=yellow]{and different bearing models}{Added based on reviewer comment 1.2}, is recommended to fully validate the method's usefulness for diagnosis.

\section{Acknowledgment}
\ifx\templateType\templateIEEE
This research was supported by FedDev Ontario project 814996, Natural Sciences and Engineering Research Council of Canada (NSERC) Create project CREAT-482038-2016 and D\&V-NSERC Alliance project ALLRP-549016-2019. \fi
The authors would like to kindly thank FCA US LLC and the EECOMOBILITY team for their collaboration.
We sincerely thank Elliot Huangfu, Ehsan Majma, Edgar Vazquez, and Jonathon Wong for their work on the BSG project.

\bibliographystyle{IEEEtran.bst}

\bibliography{pub.bib}

\end{document}